\documentclass[draft]{agujournal2019}
\usepackage{url} 
\usepackage{lineno}
\usepackage[inline]{trackchanges} 
\usepackage{soul}
\draftfalse

\journalname{JGR: Space Physics}

\begin{document}

\title{A Localized Burst of Relativistic Electrons in Earth's Plasma Sheet: Low- and High-Altitude Signatures During a Substorm}

\authors{M. Shumko\affil{1}, D.L. Turner\affil{1}, A.Y. Ukhorskiy\affil{1}, I.J. Cohen\affil{1}, G.K. Stephens\affil{1}, A. Artemyev\affil{2}, X. Zhang\affil{3}, C. Wilkins\affil{2}, E. Tsai\affil{2}, C. Gabrielse\affil{4}, S. Raptis\affil{1}, M. Sitnov\affil{1} and V. Angelopoulos\affil{2}}

\affiliation{1}{Johns Hopkins University Applied Physics Laboratory, Laurel, MD, USA}
\affiliation{2}{Department of Earth, Planetary, and Space Sciences, University of California Los Angeles, Los Angeles, CA, USA}
\affiliation{3}{Department of Physics at University of Texas at Dallas, Texas, USA}
\affiliation{4}{The Aerospace Corporation, El Segundo, CA, USA}

\correspondingauthor{M. Shumko}{mike.shumko@jhuapl.edu}

\begin{keypoints}
\item Earth's plasma sheet near $33 \ \mathrm{R_E}$ rapidly accelerated a 3 MeV electron burst and scattered it into the atmosphere
\item This burst was not observed at $17 \ \mathrm{R_E}$ but was concurrent with bursty bulk flows and dipolarization flux bundles at that location
\item While the electron acceleration mechanism was not directly observed, we deduce that it was extremely efficient, rapid, and localized
\end{keypoints}

\begin{abstract}
Earth's magnetotail, and the plasma sheet embedded in it, is a highly dynamic region that is coupled to both the solar wind and to the inner magnetosphere. As a consequence of this coupling, the plasma sheet undergoes explosive energy releases in the form of substorms. One consequence of this energy release is heating of thermal electrons and acceleration of energetic (non-thermal) electrons. The upper-energy limit as well as the spatial scale size of the electron acceleration regions remain mysteries in magnetotail physics because current missions can effectively only offer us a single-point glimpse into the numerous magnetotail phenomena ranging from electron- to global-scales. These energetic electrons can provide a significant source of seed electrons for the Van Allen Radiation belts. Here we use a unique approach to study relativistic plasma sheet electron acceleration. We combine high-altitude Magnetospheric Multiscale (MMS) mission observations with low-altitude Electron Losses and Fields Investigation (ELFIN) observations, to quantify the upper-energy extent and radial scale of a burst of plasma sheet electrons that mapped to 33 Earth radii. The plasma sheet locally accelerated an intense mesoscale burst of 3 MeV electrons---far higher and more intense than the outer Van Allen radiation belt---and scattered them into the atmospheric loss cone. High-altitude observations Earthward of the burst at 17 Earth radii showed only the usual substorm activity signatures---demonstrating that this burst was 1) intense, 2) localized to the far magnetotail, and 3) likely accelerated by a very efficient and rapid mechanism.
\end{abstract}

\section*{Plain Language Summary}
\noindent Earth's magnetotail, and the plasma sheet embedded in it, is a highly dynamic region that is coupled to both the solar wind that flows outward from the sun, and Earth's inner magnetosphere which is sculpted by Earth's magnetic field. During substorms, the plasma sheet undergoes explosive reconfiguration of the magnetic fields and particles which leads to spectacular displays of the aurora. In this paper we use the complementary perspectives from NASA's Magnetospheric Multiscale (MMS) and UCLA's Electron Losses and Fields Investigation (ELFIN). We use these missions to investigate an energetic burst of electrons that was generated in the far plasma sheet and scattered into Earth's atmosphere. This burst contained relativistic electrons with energies 10x higher than in the ambient plasma sheet, and was 10x more intense than the nearby Van Allen radiation. MMS did not directly observe the burst, despite it observing the Earthward plasma sheet, hinting that the size of the burst was confined to a few Earth radii in the radial direction. Lastly, we suspect that the electron acceleration mechanisms was very efficient and rapid.

\section{Introduction}
Earth's plasma sheet is a region of hot plasma constrained between oppositely-oriented closed field lines in the magnetotail. In its quiescent state, the plasma sheet is maintained by a combination of magnetic pressure and tension forces from the lobes, closed field lines in the far magnetotail, and strong fields in the inner magnetosphere. But as the lobe pressure increases from dayside reconnection, during the process known as the Dungey cycle, the plasma sheet becomes a reservoir of energy and plasma \cite<e.g.,>{Dungey1963, Kivelson1995}. During this process the current sheet thins, eventually causing a cascade of multi-scale phenomena resulting in what is typically called a magnetospheric substorm \cite<e.g.,>{Akasofu1964, Sitnov2019}. A wide range of subsequent phenomena include reconnection exhaust jets, flux ropes (i.e., a plasmoid without a dawn-dusk magnetic field component), bursty bulk flows, injections, and dipolarization fronts \cite<e.g.,>{Angelopoulos1994, Runov2009, Gabrielse2014, Huang2019, Turner2021}. Thermal plasma sheet electrons in these structures are heated, while non-thermal (i.e., energetic) electrons are accelerated \cite<e.g.,>{Cohen2021, Oka2023}. The Earthward-flowing particles in these structures can seed the ring current and the radiation belts, as well as couple to the high latitude ionosphere via field-aligned currents \cite{Sorathia2018, Turner2016, Turner2021b}.

Here we focus our attention on transient plasma sheet electron acceleration to relativistic energies, which are about two orders of magnitude more energetic than the thermal plasma sheet electrons \cite{Retzler1969, Terasawa1976, Richardson1993, Richardson1996}. The characteristic motions of these electrons in the magnetotail make their confinement difficult: they either immediately scatter into Earth's atmosphere through loss of adiabaticity via current sheet scattering \cite<e.g.>{Sergeev1983, Wilkins2023, Zou2024} or gradient/curvature drift out of the magnetopause in a few minutes \cite{Turner2021b}. Due to their short lifetime, prior studies have suggested that these electron bursts were local accelerated in the $<80$ Earth radii ($\mathrm{R_E}$) plasma sheet during substorms \cite{Terasawa1976, Richardson1993, Richardson1996}.

The above observations were limited by their integral energy channels, and the single in-situ satellite approach lacks broader context on whether these acceleration events are local (e.g., mesoscale) or global---presently a highly contentious topic. \citeA{Gabrielse2023} points out that while MHD model results \cite{Wiltberger2015} favor mesoscale features (often defined on the order of $\mathrm{R_E}$), our current observational techniques cannot conclusively determine the scales involved \cite<e.g.,>{Nakamura2004}. One important reason for understanding their scales is quantifying how ubiquitous these phenomena are: the smaller the size and/or shorter their duration, the less likely it is to be detected by in-situ high-altitude satellites.

Here we investigate the energetic extent, intensity, size, and possible drivers of one plasma sheet burst in a case study. We combine high-altitude Magnetospheric Multiscale (MMS) \cite{Burch2016} mission observations of the plasma sheet with low-altitude Electron Losses and Fields Investigation (ELFIN) \cite{Angelopoulos2020} observations of the precipitating electrons. We use the two missions to investigate the potential acceleration mechanism(s) at play during a specific relativistic burst of electrons observed along field lines connected to the deep tail, including its upper-energy extent and radial scale size. Complementary to this case study, the \citeA{Zhang2024} parallel study statistically investigates an ensemble of these relativistic plasma sheet bursts using the ELFIN and Colorado Inner Radiation Belt Experiment \cite<CIRBE;>{Li2024} missions.

\section{Methods}
For this study, the MMS mission \cite{Burch2016} provided the high-altitude perspective, while the ELFIN mission \cite{Angelopoulos2020} provided the low-altitude perspective. Due to its high-altitude orbit, MMS is effectively stationary while the plasma sheet moves past it, thus probing the temporal evolution of the plasma sheet. Complementary to MMS, ELFIN's low-altitude orbit allows it to observe the radiation belt and plasma sheet magnetic field footprints in a few minutes, effectively scanning the spatial structure of the plasma sheet. Taken together, these missions give us a spatial and temporal plasma sheet perspective.

The MMS mission is composed of four identically-instrumented satellites launched into a highly elliptical orbit on 13 March 2015 \cite{Burch2016}. The apogee during this observation was near 1 hour magnetic local time (MLT) at approximately 28 Earth radii ($\mathrm{R_E}$). Each MMS probe contains a suite of wave and plasma instruments and we use the following instruments for this study. The thermal and energetic electrons were observed by the Fast Plasma Investigation \cite<FPI;>{Pollock2016} and Fly's Eye Electron Proton Spectrometer \cite<FEEPS;>{Blake2016}, while the low- and high-frequency magnetic fields were observed by the Flux-Gate Magnetometer \cite<FGM;>{Torbert2016} and Search-Coil Magnetometer \cite<SCM;>{Russell2016, Le2016search}.

The ELFIN mission is composed of two identical 3U CubeSats launched into a polar low Earth orbit (LEO) with approximately 450 km altitude, $93^\circ$ inclination, and 90 minute orbit period \cite{Angelopoulos2020, Angelopoulos2023}. They launched on 15 September 2018 and the mission ended with their rapid orbital decay and ultimate atmospheric disintegration in September 2022 \cite{Tsai2024}. Each CubeSat was equipped with an energetic particle detector (EPD) that measured 50 keV to 7 MeV electrons in 16 logarithmically-spaced energy channels. Both CubeSats observed the pitch angle distribution by spinning. During this event, the spin period was 2.7 seconds, during which they took 16 measurements (every 177 milliseconds). Therefore, ELFIN obtained a pitch angle distribution every 1.35 seconds. As we describe in \ref{pad_calculation}, for each spin we calculated the strength of the scattering mechanism via the ratio of the precipitating to trapped fluxes ($j_{||}/j_\perp$). A ratio $j_{||}/j_\perp << 1$ indicates an empty loss cone, while ratio $j_{||}/j_\perp \sim 1$ indicates that the loss cone is full and the fluxes are locally isotropic. The energy-dependent $j_{||}/j_\perp$ ratio is an invaluable tool to distinguish different plasma populations and precipitation mechanisms \cite{Artemyev2021b, Artemyev2022, Zhang2022, Wilkins2023, Angelopoulos2023}.

To relate ELFIN and MMS, we magnetically map their locations to the magnetic equator using the data mining-based SST19 magnetic field model \cite{Stephens2019}. This novel model has been successfully used to model substorm dynamics \cite{Stephens2019,Sitnov2019b}, similar to the conditions in this study. SST19 was customized for use in this study by adding ARTEMIS \cite{Angelopoulos2010} magnetometer data to the archive and halving its radial resolution, thus extending its spatial domain to cislunar distances. We primarily use this novel model to map the ELFIN and MMS orbits to the neutral sheet during this substorm. This model does not, however, accurately predict the location and timing of mesoscale structures. Thus, model-data alignment of mesoscale features is a coincidence.

\section{Results}
We focus on a time interval spanning 17-20 UT on 17 July 2021, during which ELFIN downlinked data for three nightside passes. The geomagnetic conditions were quiet during this time interval with the exception of a small substorm during the time period of interest. The SuperMAG SME index \cite{Gjerloev2009}, a proxy of the auroral electrojet (AE) index, peaked at 337 nT during the substorm.

For context, Fig.~\ref{fig1} shows the ELFIN and MMS locations overplotted against the SST19 reconstructed magnetic field. MMS was effectively stationary during this time interval and thus was well-posed to observe transient magnetotail dynamics such as bursty bulk flows, dipolarizations, and plasma sheet flapping. ELFIN, on the other hand, completed multiple LEO orbits during this time period and collected data for three passes that are labeled C1--C3 in Fig.~\ref{fig1}(a) and \ref{fig1}(b): one pass at 16:57--17:02 UT, and two closely-timed passes at 19:40--19:45 and 20:00--20:06 UT. While there were more passes between 17 and 19 UT, that data was not downloaded due to ELFIN's limited downlink rate.

Figure \ref{fig1}(c1)--(c3) show the modeled $B_z$ magnetic field component on the Y-X plane for the three conjunctions. The colored dots show the MMS location, and the purple lines show the radial extent to which ELFIN maps to in the five minute pass. In other words, within a few minutes, ELFIN provides a snapshot of a large extent of the nightside magnetosphere, including the radiation belts, plasma sheet, and the polar cap. The curved ELFIN trajectory is a reflection of the non-trivial magnetic field mapping between LEO and the neutral sheet. The green star in Fig.~\ref{fig1}(c2) shows ELFIN's mapped location when it observed the plasma sheet burst at 19:41 UT---the main focus of this study. The Z-X projection of the modeled magnetic field in Fig.~\ref{fig1}(d1)--(d3) shows that MMS was just above the plasma sheet at approximately $\vec{X}_{GSM} = (-16, -2, 5)$.

\begin{figure}
\includegraphics[width=0.9\textwidth]{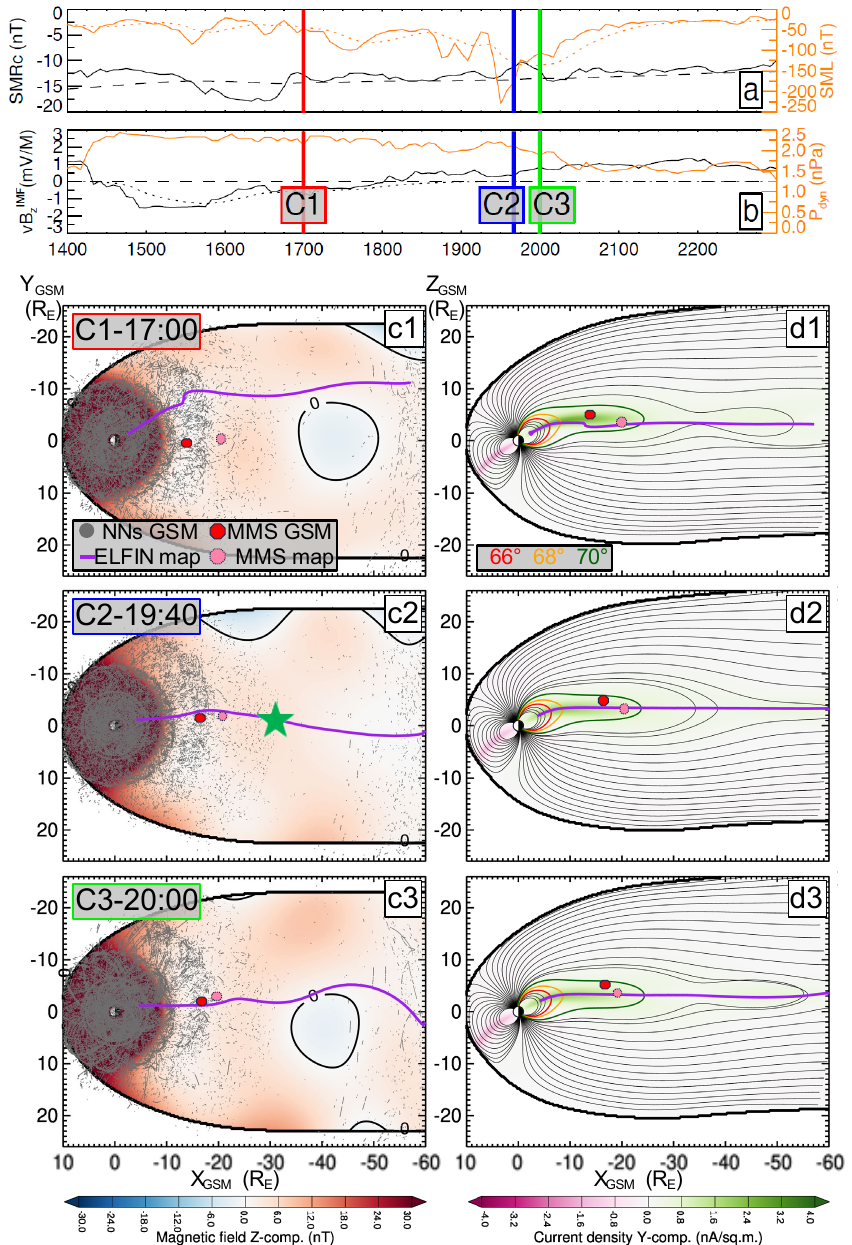}
\caption{(Caption next page.)}
\end{figure}
\addtocounter{figure}{-1}
\begin{figure} [t!]
\caption{The mapped locations of MMS and ELFIN in the SST19 reconstructed magnetic field. Panel (a) shows the pressure-corrected storm, $\textit{SMRc}$ (black line), and substorm, $\textit{SML}$ (orange line), geomagnetic indices. Their smoothed values, the inputs to SST19, are indicated by the dashed and dotted lines, respectively. Panel (b) shows the solar wind electric field parameter, $vB_z^\mathrm{IMF}$ (black line), and the dynamic pressure, $P_\mathrm{dyn}$ (orange line). The smoothed $vB_s^\mathrm{IMF}$, an SST19 input, is shown as the dotted black line. The three vertical lines, denoted by labels C1-C3, correspond to the three MMS-ELFIN conjunctions. Panels (c1)-(c3) show the modeled $B_z$ component in the Y-X plane. The grey dots indicate the locations of the nearest neighbors, that is, the magnetometer records used to fit the magnetic field description. The red and pink circles are at the actual MMS locations (in GSM coordinates), as well as the MMS location mapped to the neutral sheet. The purple lines correspond to the mapped ELFIN trajectory. Panels (d1)-(d3) show the magnetic field lines, seeded every $2^\circ$ in latitude with selected field lines highlighted, and the out-of-plane current density in the Z-X plane. The MMS locations and ELFIN trajectories are shown with the same color-coding as in Panels (c1)-(c3). The green star in Panel (c2) shows the location of the plasma sheet burst during Conjunction \#2. During Conjunction \#3 twenty minutes later, the SST19 model predicts an X-line near the site of the plasma sheet burst, indicated with the black contour, hence the mapped ELFIN trajectory is discontinuous there.}
\label{fig1}
\end{figure}

\begin{figure}
\noindent\includegraphics[width=\textwidth]{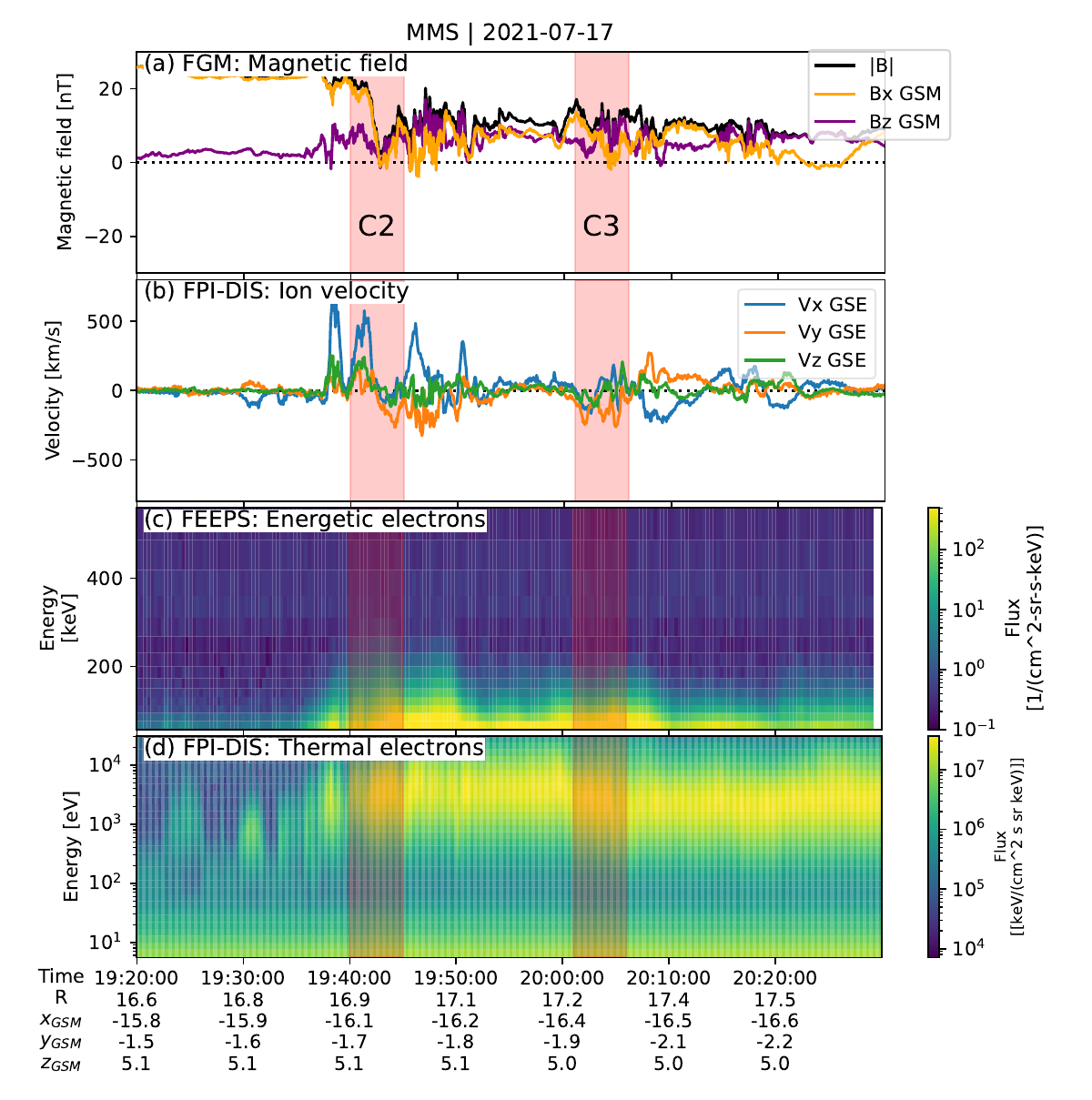}
\caption{MMS observations taken during conjunctions 2 and 3. Panel (a) shows the $B_x$ and $B_z$ magnetic field components. Panel (b) shows the ion velocity. Panels (c) and (d) show the high- and low-energy electrons. The time range of the semi-transparent red box spanning all panels corresponds to the ELFIN conjunctions 2 and 3.}
\label{fig2}
\end{figure}

Figure \ref{fig2} shows the MMS magnetic field and particle measurements during conjunctions 2 and 3. During these conjunctions, MMS was near the central plasma sheet ($|B_x| \sim 0 \ nT$) and it observed thermal electrons in the 1-10 keV range, a high energy tail of the plasma sheet electrons with energies up to $\sim 200 \ \mathrm{keV}$, bursty bulk flows up to 700 km/s, and dipolarizing flux bundles. Taken altogether, MMS observed typical substorm activity signatures.

\begin{figure}
\noindent\includegraphics[width=\textwidth]{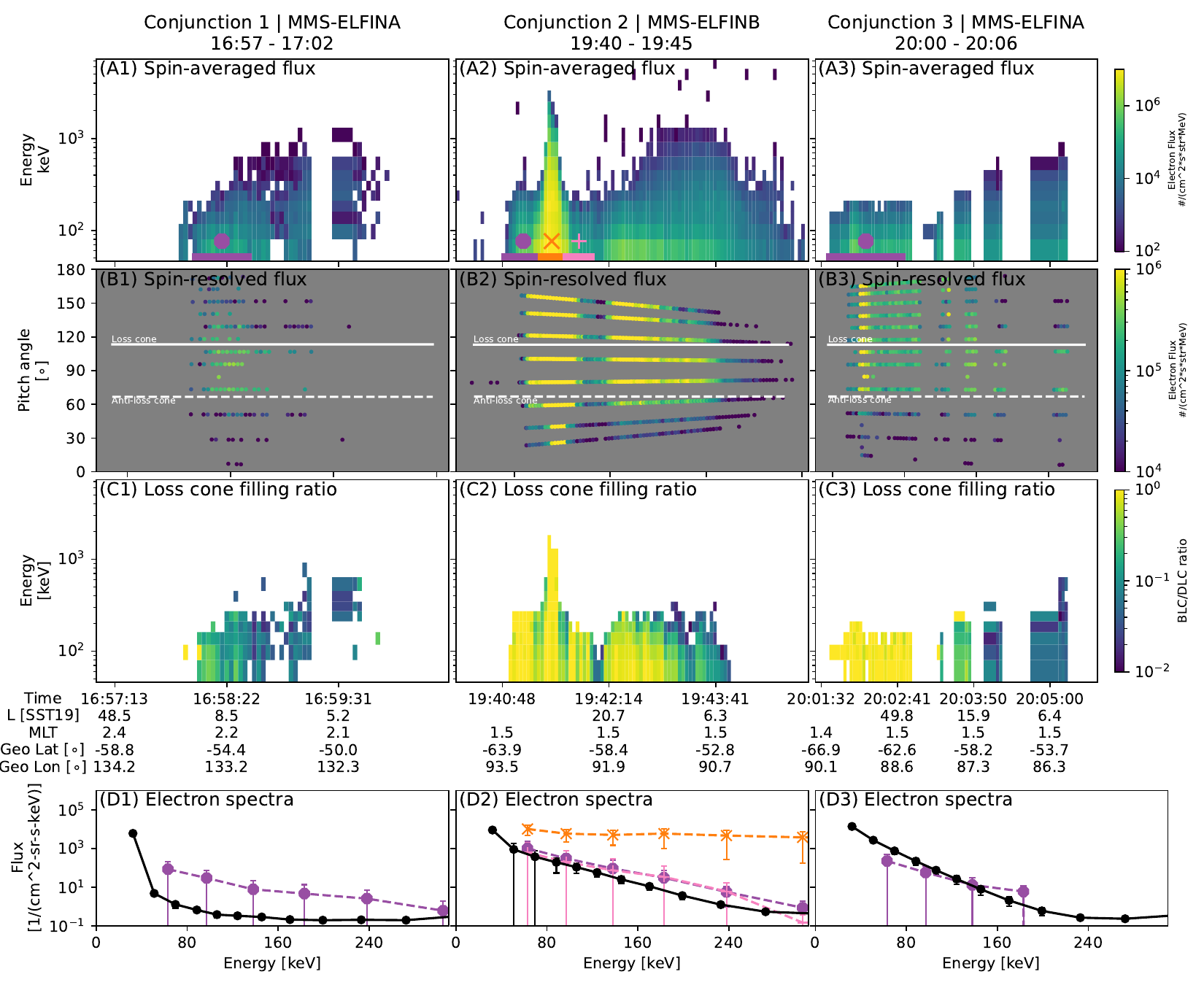}
\caption{Electron spectra during the three ELFIN-MMS conjunctions. The top row shows omnidirectional electron flux observed by ELFIN during the three conjunctions. In about 5 minutes ELFIN observes the night side magnetosphere from the radiation belts to the plasma sheet. Only fluxes that correspond to $>5$ counts are plotted. This count threshold, together with missing data, corresponds to the empty fluxes in the panels. The second row shows the 63 keV flux as a function of pitch angle and time. This shows the range of pitch angles that ELFIN sampled at the time. The third row shows the $j_{||}/j_\perp$ ratio showing the scattering efficiency. The bottom panels show the average electron spectrum from MMS-FEEPS and ELFIN. The solid black line with points shows the omnidirectional FEEPS spectrum averaged over each of the five minute conjunctions. The dashed colored lines are the ELFIN energy spectra taken from the plasma sheet region, highlighted with the corresponding colored time spans and markers in the first rows. The error bars represent the Poisson uncertainty.}
\label{fig3}
\end{figure}

Figure \ref{fig3} compares the ELFIN and MMS energy spectra during the three conjunctions. The panels A1-A3 in Fig. \ref{fig3} show the evolution of the spin-averaged plasma sheet fluxes from ELFIN's perspective. The panels B1-B3 in Fig. \ref{fig3} show the spin-resolved pitch angles that ELFIN observed, and panels C1-C3 shows the $j_{||}/j_\perp$ ratio. A ratio close to 1 indicates that the loss cone is full of precipitating electrons. The fluxes around 17 UT were fairly low and the loss cone was partially filled. But around 19:40 UT both the plasma sheet and the outer radiation belt fluxes were enhanced and the loss cone mostly filled for the majority of the pass. At 19:41 ELFIN observed a very intense and latitudinally-localized burst of electrons with energies up to 3 MeV. And lastly, the following pass 15 minutes later shows that while the loss cone is still filled, the intense burst was absent. 

To link the ELFIN and MMS particle environment, panels D1-D3 in Fig. \ref{fig3} compare the high- and low-altitude spectra observed by MMS-FEEPS and ELFIN. The solid black lines show the average FEEPS spectra during each five-minute conjunction, while the color-coded dashed lines and markers show the ELFIN spectrum averaged over time periods with the corresponding horizontal colored lines. Up until 19:40, MMS spent most of its time in the lobes, thus the FEEPS spectrum is representative of the deep tail and is systematically lower than ELFIN. The error bars represent the standard deviation assuming Poisson statistics.

In general, the high- and low-altitude spectra agree to within the Poisson uncertainty when MMS observed the plasma sheet during Conjunctions 2 and 3. The stark exception is the spectra taken during the energetic plasma sheet burst observed at 19:41 UT by ELFIN during Conjunction 2. For most energies above $\sim 100$ keV, the burst flux was orders of magnitude higher than what MMS observed, and extended to 3 MeV as Fig. S1 shows in more detail. This appears to be a burst of electrons localized to field lines threading ELFIN and not MMS, with fluxes on field lines that map Earthward and tailward of the burst in good agreement between MMS and ELFIN.

\section{Discussion and Conclusions}

While MMS observed transient magnetotail dynamics over three hours, ELFIN rapidly scanned the entire plasma sheet three times: once at the start of the three hour interval, and twice within fifteen minutes near the end of the interval. During Conjunction 2, the low-altitude fluxes that mapped Earthward and tailward of the plasma sheet burst (the purple and pink horizontal lines in Fig. \ref{fig3}(A2)), agree within Poisson uncertainty with the high-altitude plasma sheet spectrum. Therefore, ELFIN was magnetically connected to the plasma sheet on either side of the burst \cite{Artemyev2022, Artemyev2023}. This spectral comparison confirms that 1) the burst originated from the plasma sheet, and 2) MMS was not at the acceleration site. From ELFIN-B we also ascertain that the energy spectrum of the burst that was observed at 19:41 UT was much harder and more intense than the outer radiation belt. Furthermore, the upper-energy extent reached 3 MeV---well within the relativistic regime. As we demonstrate in \ref{saturation}, ELFIN underestimated the absolute fluxes due to saturation. Our first takeaway is that ELFIN observed a burst of plasma sheet electrons with fluxes at least an order of magnitude above the outer Van Allen radiation belt. Since the burst fluxes were higher than the radiation belt, this excludes the possibility of a sudden magnetic field reconfiguration that mapped the burst to the outer radiation belt \cite{Chu2015}.   

Since MMS did not observe the burst, we presume that ELFIN observed the spatial extent of the burst. We use the SST19 magnetic field model to map the Earthward and tailward boundaries of the burst from LEO to the plasma sheet. During the approximately ten seconds ELFIN scanned the burst, its equatorial footprint moved from 36 to 33 $\mathrm{R_E}$. Thus, the burst size in primarily the radial direction is about $3 \ \mathrm{R_E}$. Considering model uncertainty, the exact size of the burst is unknown, but we conclude that the model maps the burst beyond MMS. Indeed, $3 \ \mathrm{R_E}$ rules out a global instability and is instead similar to many mesoscale phenomena observed in the plasma sheet, suggesting a potential connection \cite<e.g.,>{Wiltberger2015, Sorathia2021, Ukhorskiy2022, Gabrielse2023}.

Now we consider the possible mechanisms capable of accelerating $<$200 keV plasma sheet electrons to at least 2 MeV---a 10$\times$ increase---and then promptly scattering them into the loss cone. Since there were no high-altitude in-situ observations at the burst acceleration site, we must form our conjecture primarily based on ELFIN observations and partially on MMS observations of the plasma sheet. We begin by considering if adiabatic acceleration due to betatron and/or Fermi processes could have accelerated those electrons \cite<>[and references within]{Oka2023, Fu2020}.

Betatron acceleration preserves the first adiabatic invariant when an electron is transported into a region of increased magnetic field strength, such as a dipolarization (i.e., a magnetic island) \cite{Turner2016, Gabrielse2017, Ukhorskiy2018, Eshetu2019, Ukhorskiy2022}. For this to be plausible three criteria must be satisfied: 1) the transient magnetic ``island", which is responsible for trapping and transporting particles, must must have a size of at least $3 \ \mathrm{R_E}$ to trap this burst; 2) these electrons must stay trapped while the magnetic field strength increases 10$\times$ from $\sim$5 nT (a conservative assumption based on MMS observations) to $\sim$50 nT; and 3) these electrons, now with a significant perpendicular velocity, must impulsively scatter into the loss cone. We stress that in the third step the electron transport in pitch angle must be substantial since betatron predominately accelerates the component of the kinetic energy that is perpendicular to the background magnetic field. The electron detrapping could occur as soon as the electrons escape the dipolarized magnetic field into the ambient plasma sheet magnetic field, whose field lines are typically curved enough to scatter $>$60 keV electrons \cite{Artemyev2022, Artemyev2023}.

Fermi acceleration conserves the second adiabatic invariant as the particle's path parallel to the background magnetic field shortens \cite{Fermi1949, Northrop1963, Drake2006, Mozer2016, Arnold2021}. One possibility is a two-step process: 1) while escaping the reconnection site in the exhaust, the plasma sheet electrons became confined in a plasmoid on field lines that progressively shrink to a length 10$\times$ shorter than the field lines on which the electrons were on before, and 2) the trapped electrons must impulsively scatter, similar to the betatron case. This is the scenario described in \citeA{Richardson1993} and \citeA{Richardson1996}

While both Fermi and betatron acceleration mechanisms are plausible explanations of the burst, they require multi-step processes. Thus, we investigate whether non-adiabatic mechanisms, such as turbulent electron-kinetic-scale magnetic and electric structures or wave-particle interactions, can efficiently accelerate and scatter these electrons \cite{Scarf1974, Cattell1986, Huang2019, Khotyaintsev2019, Usanova2022}. While turbulence can cause runaway electron acceleration to non-thermal energies, it acts on $\sim 10 \ \mathrm{R_E}$ spatial scales---much larger than the $3 \ \mathrm{R_E}$ burst size \cite{Ergun2020a, Ergun2020b}. Electrons can also be accelerated by waves such as lower hybrid waves, Langmuir waves, electrostatic solitary waves, kinetic Alfvén waves, and whistler mode waves \cite{Khotyaintsev2019, Shen2022}. However, these wave-particle interactions are highly energy-dependent, and none of the wave modes have been shown to directly accelerate electrons to relativistic energies in the magnetotail \cite<e.g.,>{Chepuri2023}. Parallel electric fields in reconnection regions can also heat electrons, but this is not a plausible mechanism in our study because these electric fields do not accelerate electrons to relativistic energies \cite<e.g.>{Egedal2012}. Considering the aforementioned adiabatic and non-adiabatic mechanisms, we cannot identify a clear culprit. This can be addressed in future studies using in-situ observations of the particle bursts made in the plasma sheet, or MHD simulations with particle transport and self-consistent wave-particle interactions.

In summary, combining high- and low-altitude observations taken by MMS and ELFIN enabled an unprecedented view of plasma sheet acceleration of relativistic electrons. The plasma sheet burst was more intense then the outer radiation belt and was localized to $3 \ \mathrm{R_E}$ in the radial direction. Although we cannot confirm the scattering mechanism, our evidence suggests that it is compact and efficient. In addition to this case study, the \citeA{Zhang2024} parallel study explores the properties of a dozen similar plasma sheet acceleration events in the multi-year ELFIN and CIRBE datasets, demonstrating that these events may not be entirely unique.

\appendix

\section{ELFIN Pitch Angle Distribution Calculation} \label{pad_calculation}
To process the ELFIN-EPD data, we first merged the counts (level 1) and flux (level 2) data products. We then masked all fluxes with corresponding count rates less than 5 counts/sector, or if the spin-period fell outside the [2.5, 3.5] second range. Next, for each energy channel we averaged the flux in spin-period time intervals, and in [0-180] degree pitch angles with $20^\circ$ steps (the approximate ELFIN-EPD field of view). The result is fluxes binned in time, pitch angle, and energy. 

To calculate the omnidirectional flux we simply averaged over the binned pitch angles. And to quantify the strength of the scattering, we calculated the ratio of the trapped to precipitating fluxes ($j_\perp/j_{||}$). For each time stamp we averaged all fluxes corresponding to pitch angles inside and between the loss- and anti loss-cone angles. The loss cone angle for particles lost at $<100$ km altitudes is a standard level 2 data product calculated using the International Geomagnetic Reference Field \cite<IGRF;>[]{Alken2021} magnetic field model and is approximately $60^\circ$ in LEO. A ratio $j_{||}/j_\perp << 1$ indicates an empty loss cone, while ratio $j_{||}/j_\perp \sim 1$ indicates that the loss cone is full and the fluxes are locally isotropic.

\section{ELFIN Electron Spectrum and Saturation} \label{saturation}
\begin{figure}
\includegraphics[width=0.9\textwidth]{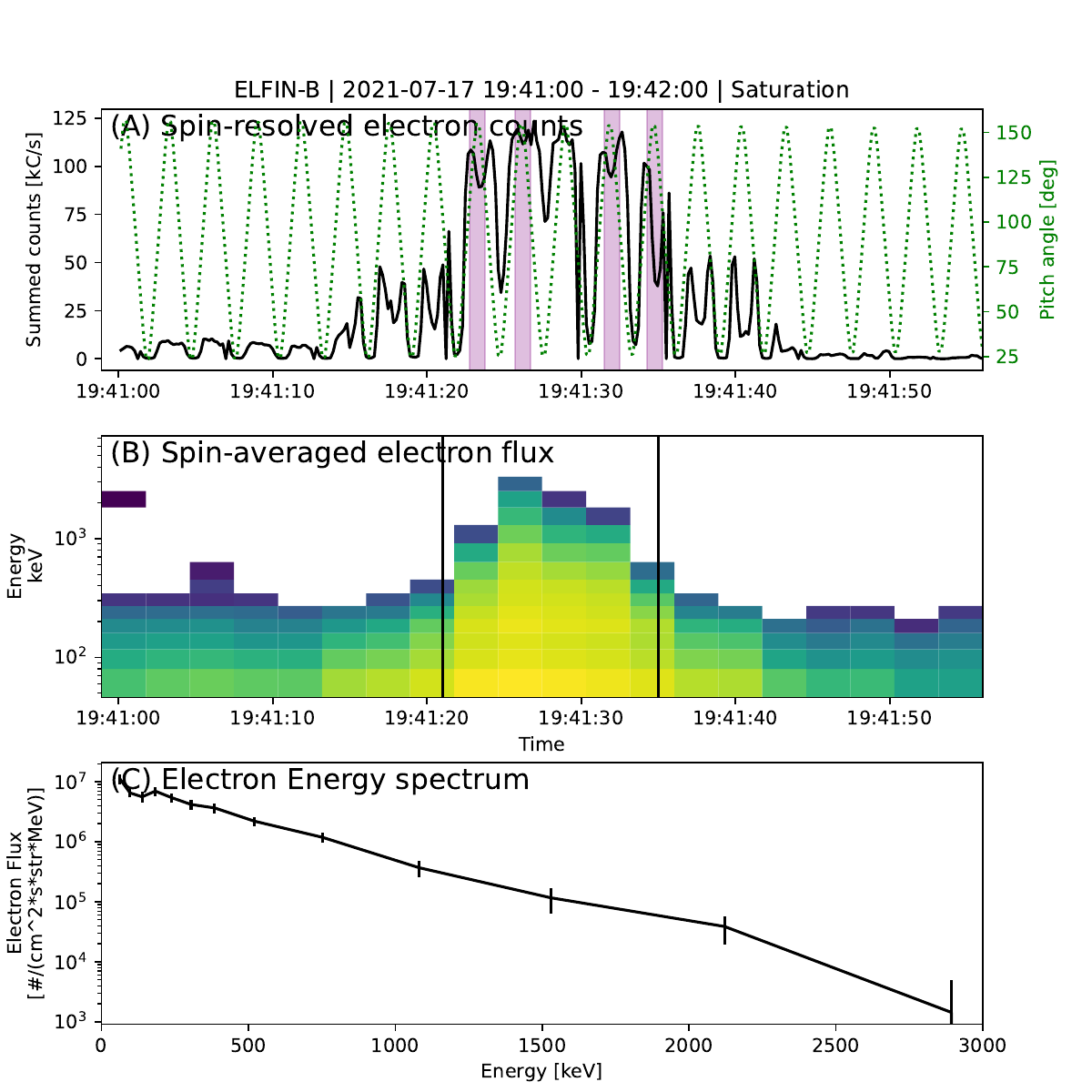}
\caption{ELFIN-EPD electron data shows saturation during the plasma sheet burst. Panel (A) shows the spin resolved counts from all energy channels in black, every 270 ms, and the oscillating dotted green line is the local pitch angle of the instrument boresight. The purple bars indicate times when EPD scanned the center of the loss cone (centered on $180^\circ$ pitch angles) and the count rates sagged, indicating that the instrument was beginning to saturate. Panel (B) is the omnidirectional electron flux in the same format as Fig. \ref{fig3}(A2) for fluxes with corresponding count rates $>3$. Panel (C) shows the electron spectrum with Poisson error bars. The spectrum was derived using the time range bounded by the vertical black lines in Panel (B).}
\label{fig_a1}
\end{figure}

ELFIN-EPD experienced signs of saturation during the burst at 19:41. This is evident for times when the total count rate from all channels exceeds $10^{5}$ counts/second. Figure \ref{fig_a1}(A) shows spin-resolved total count rates. Times when the total count rates sagged around $10^{5}$ counts/second is indicative of saturation. For reference, Panel (B) shows the spin-averaged flux in the same format as Fig. \ref{fig3}(A2). Considering the spin phase-dependent saturation (when EPD scanned through the $180^\circ$ loss cone), the true pitch angle distribution is unknown so we did not plot the $j_{||}/j_\perp$ ratio. Lastly, Panel (C) shows the energy spectrum that extends to nearly 3 MeV.

\section{Open Research}
\noindent Both the ELFIN and MMS data used for this study are publicly available at \url{https://data.elfin.ucla.edu/} and \url{https://lasp.colorado.edu/mms/sdc/public/about/browse-wrapper/}. We analyzed the MMS data using the Python library pyspedas \cite{Angelopoulos2019} version 1.5.6 that is archived at \url{https://pypi.org/project/pyspedas/} and \url{https://github.com/spedas/pyspedas}. 

\acknowledgments
We are thankful for the countless engineers, technicians, scientists, and students who made the MMS and ELFIN missions a resounding success. M. Shumko, I.J. Cohen, and S. Raptis were supported by the Magnetospheric Multiscale (MMS) mission of NASA's Science Directorate Heliophysics Division via subcontract to the Southwest Research Institute (NNG04EB99C). G.K. Stephens was supported by NASA grant 80NSSC24K0556. Lastly, this material is based upon work supported by the National Science Foundation under a Geospace Environment Modeling Grant No. (2225613).

\bibliography{refs}

\end{document}